Approaches to the economics of the
Spanish Second Republic prior to 1936

# Aproximación a los estudios sobre la economía en la Segunda República española hasta 1936

Inés Martín de Santos
**School of Business. Kendall College, Chicago**
ines.martindesantos@kendall.edu

Arturo Martín Vega
**Universidad Carlos III de Madrid**
arturom@bib.uc3m.es

## Resumen

Los datos macroeconómicos sobre la economía española durante la Segunda República no son del todo bien conocidos, y la interpretación de los hechos históricos, a partir de las cifras obtenidas, resulta divergente y tendenciosa. Los problemas sociales, derivados fundamentalmente de las profundas desigualdades económicas, se intentaron paliar con leyes apresuradas que en muchos casos supusieron declaraciones de buenas intenciones. España arrastraba el declive económico internacional que comenzó a sentirse a finales de la dictadura del

## Abstract

Macroeconomic data on the Spanish economy during the Second Republic is not accurate, the interpretation of historical events from the figures obtained is divergent and misleading. Hasty laws were enacted in attempts to resolve social problems arising mainly from deep economic inequalities, but they were often nothing more than declarations of good intentions. Spain suffered in the aftermath of the international economic downturn as it began to be felt at the end of the dictatorship of General Primo de Rivera. Eco-general Primo de Rivera. La política económica se desarrolló al amparo de la Constitución y, a pesar de las discrepancias entre el primer y el segundo bienio, fue continuista respecto a la etapa anterior y, en general, si no justa a veces, sirvió al menos para evitar la desestabilización del sistema financiero. En todo caso, resultó insuficiente para realizarse plenamente, debido sobre todo a los abundantes cambios gubernamentales, y mediatizada por una crisis social de mayor transcendencia que relegó los problemas económicos a un segundo plano.

nomic policy was developed under the Constitution, but, despite the differences between the first and second biennium, there was a tendency to maintain the guidelines from the previous stage and in general, sometimes unfairly, it aimed at least to avoid the destabilization of the financial system. Nonetheless, it ultimately failed to achieve its goals, mainly because of the frequent changes of government mediated by a social crisis of greater significance that had relegated economic issues into the background.

## 1. Introducción

Teniendo en cuenta la envergadura del tema que nos ocupa, conviene seleccionar los estudios más importantes sobre la economía española durante la Segunda República hasta el advenimiento de la guerra civil. Francisco Comín Comín[1] marcó este sendero hace más de treinta años con la publicación de una guía bibliográfica. Gloria Núñez Pérez[2] continuó el ejemplo de Comín ampliando la recopilación de obras no solo a los asuntos económicos. Desde entonces han aparecido nuevas contribuciones pero no siempre con idénticas interpretaciones.

Como señala Comín en el citado artículo[3], la avalancha de publicaciones sobre este tema se produce a partir de 1975, y añade: «La razón es obvia» (probablemente se refiera a las dificultades de hacerlo durante la dictadura franquista).

Sin embargo, la afirmación de Comín debe ser contrastada con estudios de carácter bibliométrico ya que, al menos en cuanto a publicaciones unitarias o monografías se refiere, no hemos observado un elevado incremento de las mismas ni tras 1975 ni con motivo del cincuentenario de la proclamación de la Segunda República. Los primeros estudios salieron a partir de 1973, y el año en el que más libros aparecieron sobre esta materia fue 2006. En la actualidad queda mucho por hacer pero no tanto como para que la labor pendiente sea de cien años como ha afirmado Paul Preston[4].

Las fuentes de autoría unipersonal más relevantes sobre la historia económica de este período son básicamente, entre otras, las monografías de Ricardo Calle[5], Juan Hernández Andreu[6], Jordi Palafox Gamir[7] y Ángel Viñas Martín[8]. Además de aportar abundante docu-mentación, realizan una objetiva interpretación de los datos. Un período más extenso abarca el más reciente y renovador manual de Albert Carreras y Xavier Tafunell[9], que analiza el nivel de vida de la sociedad española desde la perspectiva comparativa respecto a los países europeos más desarrollados.

---

1. F. Comín, "Una guía bibliográfica para el estudio de la economía en la Segunda República española", *Revista de Estudios Políticos,* 31-32, 1983, 313-334.
2. G. Núñez Pérez, *Bibliografía comentada sobre la Segunda República española: obras publicadas entre 1940 y 1992*, Madrid, 1993.
3. Comín, "Una guía bibliográfica…" *op. cit.*, 315.
4. Apud M.L. de Prado Herrera, "La historiografía de la guerra civil y del primer franquismo: reflexiones y nuevos planteamientos en el setenta aniversario", *Studia Historica. Historia Contemporánea*, 25, 2007, 321.
5. R. Calle Saiz, *La hacienda pública en la II República española*, Madrid, 1981, 2 vols.
6. J. Hernández Andreu, *España y la crisis de 1929*, Madrid, 1986.
7. J. Palafox, "La economía", en S. Payne y J. Tussel (dirs.). *La guerra civil. Una nueva visión del conflicto que dividió España*, Madrid, 1996, 195-265.
8. Á. Viñas, *El oro de Moscú: alfa y omega de un mito franquista*, Barcelona, 1979.

Ante la inmensa bibliografía sobre este período, muchos estudios han acotado campos parciales y se han centrado en las economías regionales, sobre todo a partir de los años noventa[10]. Son menos, en cambio, los de carácter microeconómico, que resultan tan importantes como los de tipo macroeconómico, para conocer el grado de bienestar en las economías domésticas.

Algunos científicos han restringido el ámbito de trabajo a las personas. En esta línea, Indalecio Prieto es probablemente una de las personalidades más tratadas, aunque valorado de manera positiva como hace Juan Velarde[11] o negativa como interpreta Francisco Olaya[12]. No menos relevante es el caso del doctor Juan Negrín, presidente del gobierno republicano entre 1937 y 1939 (posteriormente hasta 1945 en el exilio), ministro de Hacienda en el gobierno de Largo Caballero y máximo antagonista del general Franco según la exhaustiva biografía de Moradiellos[13]. En el caso de Negrín, la fama le ha resultado mayoritariamente adversa hasta su apropiada reivindicación por Miralles[14].

En todo caso, resulta una tarea difícil escudriñar y consensuar una única versión sobre el pensamiento de los economistas españoles durante la Segunda República española desde la perspectiva historiográfica, debido por una parte a la bibliografía que falta por revisar, con interpretaciones no siempre coincidentes; y por otra parte, a la todavía inacabada exploración de muchos archivos, acontecimientos, datos e ideas por descubrir e interpretar como afirma Enrique Fuentes Quintana[15].

---

9. A. Carreras y X. Tafunell, *Historia Económica de la España Contemporánea (1789-2009)*, Barcelona, 2010.

10. Cfr J.R. Cuadrado Roura, "El desarrollo de los estudios de economía regional en España", *Revista de Estudios Regionales*, 75, 32.

11. J. Velarde Fuertes, "Indalecio Prieto en Hacienda", *Estudia historica. Historia Contemporánea*, 1, 1983, 53-66.

12. F. Olaya Morales. *Negrín, Prieto y el patrimonio español*, Madrid, 1996.

13. E. Moradiellos García, *Don Juan Negrín*, Barcelona, 2006.

14. R. Miralles, *Juan Negrín: la República en guerra*, Madrid, 2003. El profesor Juan Negrín, tardíamente incorporado a la política, fue, a nuestro juicio, fundamentalmente un tecnócrata y por más que tuviera una acertada política económica que podríamos denominar de tinte keynesiano en algunos aspectos (en aquellos tiempos las ideas de Keynes todavía no se conocían en España), algunas de sus decisiones como el famoso oro de Moscú le han hecho pasar por una de las personas más impopulares de entonces, incluso entre sus partidarios. Sobre el tema de las reservas de oro del Banco de España, consúltese el ecuánime y clarificador estudio de P. Martín Aceña, *El oro de Moscú y el oro de Berlín*, Madrid, 2001.

15. E. Fuentes Quintana (dir.) y F. Comín (coord.), *Economía y economistas españoles en la guerra civil*. Madrid, 2008, 17.

Uno de los obstáculos con los que se enfrenta la historiografía es el de la interpretación de los hechos, sobre todo si los investigadores no vacían y analizan de manera exhaustiva las fuentes primarias, o si actúan influidos por cuestiones ideológicas o intereses particulares como comenta el profesor Viñas[16].

Esta parcialidad se hace aún más evidente cuando se trata de analizar conflictos bélicos. Sobre el periplo siguiente, acerca de la guerra civil española, por ejemplo, han aparecido infinidad de versiones: la perspectiva de la derecha (Ricardo de la Cierva,…) que manipula tanto los orígenes como el curso del conflicto, la perspectiva de la izquierda (Gabriel Jackson,…) que analiza la contienda tomando como referencia dos fuerzas organizadas, y otra vía en la extrema izquierda que presenta una visión entre un ejército regular y un pueblo obligado a improvisar su defensa a la vez que llevaba a cabo una revolución social (Vernon Richards,...).

El análisis de las decisiones económicas durante la Segunda República española es algo complicado de hacer porque está muy mediatizado por intereses muy variados, y hay que tener en cuenta que durante un período de ocho años (14 de abril de 1931 – 1 de abril de 1939) hubo nada menos que veintiséis gobiernos de coalición. Lo sorprendente es ver cómo en un período de tiempo tan corto pudo haber tanta inquietud por mejorar la sociedad.

En este trabajo sólo contemplamos algunos aspectos generales de la economía española hasta el comienzo de la contienda bélica y, dado que a menudo circula el tópico de que todas las guerras son económicas o, mejor dicho, originadas por móviles económicos, ¿Fueron determinantes las circunstancias económicas para el estallido de la guerra civil? Esta es la pregunta a la que nuestras pesquisas pretenden responder.

El período seleccionado necesita el análisis complementario del momento posterior. Durante el período bélico, la actividad productiva está determinada por una economía de guerra en ambos bandos, ampliamente estudiada y plasmada entre otros, en la ejemplar compilación preparada por Enrique Fuentes Quintana y Francisco Comín Comín[17] y sobre todo, la destacable labor de José Ángel Sánchez Asiaín[18]. Ante un final de la guerra previsiblemente corto, al menos desde la perspectiva del bando sublevado, la República sólo pudo hacer frente a los sublevados mientras se mantuvo su capacidad financiera. En su exhaustivo estudio, Sánchez Asiaín llega a analizar las reservas republicanas incluso hasta las actividades de su gobierno en el exilio.

---

16. Á. Viñas, *Salamanca 1936*, Barcelona, 2014, 7.
17. E. Fuentes Quintana y F. Comín Comín, *Economía y economistas españoles en la guerra civil*, Barcelona, 2008.
18. Cfr. J.Á. Sánchez Asiaín, *La financiación de la guerra civil española*, Barcelona, 2012.

## 2. La legislación durante la Segunda República

Partimos de la legislación porque debe ser la base del comportamiento humano en las civilizaciones modernas. En nuestro campo de estudio, las leyes, las costumbres y las relaciones sociales entre los distintos elementos del proceso de producción y distribución contribuyen a formar el armazón de cualquier sistema económico.

La Segunda República se opuso tenazmente desde sus inicios a la herencia derivada de un sistema oligárquico y que acababa de pasar por una dictadura apoyada por el rey, y que a lo más que aspiró, si es que aspiró a algo en el plano económico, fue a continuar un proceso de modernización autoritaria. Con este propósito, entre los cometidos más urgentes se impuso el de elaborar una Constitución.

Las nuevas Cortes Constituyentes acordaron la formación de una Comisión Jurídica Asesora para la preparación del texto, y esta, a su vez, designó una subcomisión de especialistas en derecho político.

En poco más de medio año, desde la proclamación de la Segunda República, se promulgó la *Constitución* el 9 de diciembre de 1931.

Los contenidos sobre materia económica, que figuran en los artículos 14, 26, 33, 44 y 47, dicen:

*Artículo 14: Son de exclusiva competencia del Estado español la legislación y la ejecución directa en las materias siguientes:*

*8ª Régimen arancelario, Tratados de Comercio, Aduanas y libre circulación de las mercancías.*

*Artículo 26: Todas las confesiones serán consideradas como Asociaciones sometidas a una ley especial.*

*El Estado, las regiones, las provincias y los Municipios, no mantendrán, favorecerán, ni auxiliarán económicamente a las Iglesias, Asociaciones e Instituciones religiosas.*
*Una ley especial regulará la total extinción, en un plazo máximo de dos años, del presupuesto del Clero.*

*Artículo 33: Toda persona es libre de elegir profesión. Se reconoce la libertad de industria y comercio, salvo las limitaciones que, por motivos económicos y sociales de interés general, impongan las leyes.*

*Artículo 44: Toda riqueza del país, sea quien fuere su dueño, está subordinada a los intereses de la economía nacional y afecta al sostenimiento de las cargas públicas, con arreglo a la Constitución y a las leyes. La propiedad de toda clase de bienes podrá ser objeto de expropiación forzosa por causa de utilidad social mediante adecuada indemnización, a menos que disponga otra cosa una ley aprobada por los votos de la mayoría absoluta de las Cortes. Con los mismos requisitos la propiedad podrá ser socializada.*

*Artículo 47: La República protegerá al campesino y a este fin legislará, entre otras materias, sobre el patrimonio familiar inembargable y exento de toda clase de impuestos, crédito agrícola, indemnización por pérdida de las cosechas, cooperativas de producción y consumo, cajas de previsión, escuelas prácticas de agricultura y granjas de experimentación agropecuarias [sic], obras para riego y vías rurales de comunicación. La República protegerá en términos equivalentes a los pescadores.*

\* \* \*

De este extracto se pueden obtener algunas conclusiones, como la trascendencia que se concedía al sector primario para el desarrollo económico del país, la erradicación de privilegios eclesiásticos que partían del Concordato de 1851, y la prevalencia del interés general o público sobre el privado.

El texto constitucional fue duramente criticado por personas de la categoría de Alcalá Zamora, Ortega y Gasset o Unamuno por unas razones que ahora no vienen al caso, pero desde la perspectiva económica se puede percibir su espíritu claramente socializante e intervencionista, en línea con las tendencias que ya se advertían en la Constitución de Weimar.

La *Constitución* reflejó la amalgama de muchas divergencias dentro del republicanismo. Fue una Constitución modernizante, pero en muchas ocasiones no satisfizo las aspiraciones de los republicanos moderados, descendientes del liberalismo decimonónico y considerados burgueses. Siendo como fue una admirable Constitución, era sin embargo, como afirmó el profesor Julio Aróstegui[19], una Constitución de izquierdas, una Constitución impuesta por un grupo a otro.

La *Constitución*, así como muchas otras leyes y disposiciones jurídicas derivadas de ella, intentó mejorar el bienestar social. Estas ideas, en algunos casos, se adelantaron a su tiempo (edad mínima para el trabajo, derecho de las mujeres a opositar,…) pero también se vieron sometidos a una contundente oposición por parte de los intereses económicos y sociales conservadores. No en vano se trataba de una renovación jurídica contraria a privilegios añejos, y dominada por el deseo de avanzar en términos de justicia social. Es cierto que la *Constitución* fue terminada de manera algo precipitada, pero, a nuestro juicio, uno de sus defectos fue el de propiciar cambios bruscos en una sociedad dividida y poco habituada a ceder derechos adquiridos.

\* \* \*

El período de mayor intensidad legislativa tuvo lugar durante el primer bienio republicano. He aquí algunas medidas económicas relevantes:

---

19. J. Aróstegui, "La guerra civil española. El nacimiento de la Segunda República", *La aventura de la historia*, Madrid, 1997.

**1931**
Este fue el año en el que más se legisló sobre economía. El 1 de julio apareció regulada la jornada de 8 horas y el sistema de libertad subsidiada o seguro de desempleo voluntario que «obtuvo unos resultados muy modestos en cuanto al número de afiliados se refiere»[20]. El 26 de noviembre se puso al día la Ley de ordenación bancaria, con una clara intención intervencionista, y el Proyecto sobre la reforma agraria.

**1932**
En 1932 se implantó el Plan de Riegos (13 de abril) con las obras en cinco pantanos y un canal, y se impulsó la aplicación de la reforma agraria (*Gaceta de Madrid*, 21 de septiembre) tras la sublevación del general José Sanjurjo.
El 20 de diciembre, a instancia del ministro de Hacienda, Jaume Carner, se promulgó la Ley de Contribución General sobre la Renta, una auténtica primicia. El pago de impuestos se calculaba de acuerdo con el sistema de estimación de bases. La base imponible se determinaba de acuerdo con los signos externos de la renta gastada y los signos externos de la renta percibida, «propiedades bien inútiles cuando el fraude era la norma general»[21].

**1933**
El 13 de diciembre se creó el Instituto de Reforma Agraria y se implantó el Plan Nacional de Obras Hidráulicas.

**1934**
Bajo un gobierno republicano de derechas y apoyado por la CEDA, se modificó la Ley de Reforma Agraria. Quedó anulada la mayor parte de las expropiaciones de uso. El ministro de Agricultura Nicasio Velayos Velayos llegó a anular incluso la Ley de Yunteros defendida por su antecesor en el cargo, el cedista Manuel Jiménez Fernández, que amparaba a los trabajadores extremeños del campo.

**1935**
Se puso en práctica la contrarreforma agraria. La orientación política republicana había dado un giro espectacular.
Ley de la previsión contra el paro (25 de junio) o Ley Salmón. Pretendía mitigar el paro mediante la inversión pública en la construcción de viviendas para clases medias. Su éxito fue relativo porque quedó circunscrita fundamentalmente a Barcelona y Madrid.

---

20. S. Espuelas Barroso, "La creación del seguro de desempleo en la II República. Un análisis de su impacto y de por qué fue voluntario" en *IX Congreso Internacional de la Asociación Española de Historia Económica*, 2008, 30.
21. E. Fuentes Quintana, *La reforma fiscal y los problemas de la hacienda pública española*, Madrid, 1990, 120.

En el mes de julio, Joaquín Chapaprieta, como ya lo hiciera Calvo Sotelo en 1930, aplicó una política restrictiva, propuso la reducción del gasto público (que entonces se denominaba *política de economías*, y hoy llamaríamos de *recortes*), una medida, a juzgar por el análisis de Jordi Palafox[22], desafortunada porque sus beneficios superaban a los producidos por la inversión privada, aun cuando anteriormente el sector primario no fuera el más potenciado.

En resumen, la preocupación de los políticos por solventar los problemas económicos y, en consecuencia, sociales, fue intensa. Ahora bien, cada tendencia lo hizo a su modo. En cualquier caso, por elemental conocimiento de Teoría Económica, tanto los gobiernos republicanos conservadores como progresistas incurrieron en contradicciones, al menos cuando pretendían equilibrar el Presupuesto y, a la vez, evitar el desempleo y fomentar el crecimiento económico.

## 3. La herencia del pasado y el mapa económico del país

A veces, para comprender la actitud de las personas y de los grupos de poder en un determinado momento, es necesario mirar hacia atrás y analizar los hechos de manera diacrónica.

En general, la economía de cada país es una historia de altibajos, de épocas decadentes y esplendorosas, lo que suele conocerse como historia de los ciclos. La nuestra no escapa a tal regla. Ahora bien, además de aquellos momentos de vacas flacas y gordas, es aconsejable analizar el estado de una economía comparándolo con el estado de otras.

En este sentido, la antes citada obra de Albert Carreras y Xavier Tafunell es certera al considerar que, aun reconociendo que la recuperación de la economía española ha sido notable en los últimos siglos, no se percibe del todo una alta convergencia con los países europeos más desarrollados.

Desde un punto de vista retrospectivo, España no supo beneficiarse de los momentos ventajosos en el pasado. En dos ocasiones, al menos, perdió la oportunidad de convertirse en una de las mayores potencias económicas del mundo. La primera a raíz del descubrimiento de América. La segunda vez tuvo lugar durante el período de entreguerras; lejos de aprovechar su neutralidad en la coyuntura bélica y de reinvertir en las propias empresas los beneficios obtenidos a partir de la primera guerra mundial, los excedentes se invirtieron en fastuosas construcciones y otros lujos exagerados.

Esto se produjo principalmente en las zonas industrializadas (País Vasco y Cataluña). Se puede afirmar que la balanza de pagos nunca llegó a ser positiva el resto del tiempo. Además España siguió siendo un país de enormes diferencias entre pobres y ricos.

La historia económica de España, *grosso modo*, se ha caracterizado la mayor parte de las veces por un sistema que pudiéramos definir, de manera metafórica, introvertido y poco abierto al exterior. El proteccionismo predominante en nuestra economía arranca, según Pe-dro Fraile[23], en la era moderna desde el momento de la Restauración monárquica, pero lo cierto es que esta tendencia ya se vislumbraba desde el comienzo de la unificación española en el siglo XVI.

---

22. J. Palafox, *Atraso económico y democracia: la Segunda República y la Economía Española: 1892-1936*, Barcelona, 1991.

Aunque el citado Fraile establece y diferencia períodos, y aprecia diferencias de matiz, la opinión más generalizada acerca de las decisiones de política económica durante la Segunda República es la que sostiene la práctica de un mayor o menor continuismo intervencionista tanto antes como después de dicho interregno.

Los siete largos años de dictadura militar al amparo de la monarquía, que precedieron a la Segunda República, acentuaron el proteccionismo de la economía española, pero sobre todo «los problemas de la Hacienda española surgieron con la gestión de Calvo Sotelo»[24].

Sin embargo, con el cambio de sistema político, la desconfianza de los inversores cundió y se produjo una notoria evasión de capitales al exterior. Esta actitud nada patriótica hizo decir a Maura que si él gobernara haría fusilar a los pesimistas[25].

Otro escollo se sumó a nuestro desarrollo económico en el siglo pasado. Gran parte de la economía mundial desde el primer tercio hasta el último tercio estuvo mediatizada por la crisis norteamericana desatada en 1929. El descalabro de la economía de Estados Unidos se extendió al exterior. En Europa comenzó a sentirse en 1931 con la quiebra del Credit Anstalt Bank de Viena[26], y su efecto dominó en el resto del continente. Aún hoy día es difícil calibrar la incidencia del crack del 29 en España. Las cifras bailan a gusto de algunos historiadores y la importancia de este fenómeno, sostenida por Hernández Andreu[27] ha sido en buena parte desmentida por Joseph Harrison[28].

Al parecer, en nuestro país esta influencia negativa no fue tan pronunciada debido al modelo de crecimiento hacia el interior y al autoconsumo, que favorecieron una menor dependencia de los mercados internacionales[29]. Es cierto que la crisis internacional se dejó notar en las exportaciones, sobre todo del sector agrícola, pero ello no supuso gravísimos perjuicios para nuestra agricultura. El problema no era tanto la salida de los productos como el excedente de mano de obra atrincherada (valga la metáfora) en las zonas rurales que no se sabía dónde colocar.

---

23. P. Fraile Balbín, "La intervención económica durante la Segunda República", en J. Velarde Fuertes (coord.), *1900-2000 Historia de un esfuerzo colectivo: cómo España superó el pesimismo y la pobreza*, Madrid, 2000, vol. I, 403-455.

24. F. Comín, "Hacienda y Economía en la España contemporánea (1800-1936*)*", Madrid, 1988, II, *La Hacienda transicional*, Madrid, 1988, II, 1021.

25. M.A. González Muñiz, *Problemas de la Segunda República,* Madrid, 1974, 15.

26. González Muñiz, *Problemas… op. cit*., 7.

27. Hernández Andreu, *España… op. cit*.

28. J. Harrison, "Hernández Andreu y la crisis de 1929", *Revista de Historia Económica*, 1, 1987, 133-138.

29. Cfr. F. Comín, "Política y economía: los factores determinantes de la crisis económica durante la Segunda República (1931-1936)". *Historia y política: ideas, procesos y movimientos sociales*, 26, 2011, 47-79.

Si echamos mano de las estadísticas[30], el mayor receso en la exportación se produjo en 1930, en tiempos de la dictadura del general Primo de Rivera, debido fundamentalmente a la crisis exterior. Sin embargo no se produjo ningún desastre económico; las cosechas de cereales fueron buenas, al menos en 1931 y 1933, y el gasto social del gobierno en 1931 fue encomiable. Entre otras decisiones importantes, se crearon 6.750 nuevas escuelas[31] para afrontar un índice de analfabetismo cercano al 40% de la población.

Aunque España era un país escasamente industrializado, contaba con muchas pequeñas empresas de carácter familiar y con unas ventas de los productos obtenidos del sector primario aseguradas en el mercado nacional.

La economía española, desde una perspectiva conservadora, no parecía muy desastrosa, y no debía serlo en el plano macroeconómico. Carlos Caamaño publicó en *ABC* el 1 de enero de 1936: «En materia económica y financiera presenta el año 1935 un relieve muy acusado por la concurrencia coincidente de varios fenómenos de positivo interés. Enumerémoslos: primero, contracción extraordinaria del comercio exterior; segundo, manifestaciones características de economía dirigida; tercero, expansión y crecimiento en el sector bursátil de la confianza gubernamental; cuarto, aumento de disponibilidades monetarias; quinto, baja del interés del dinero; sexto, conversiones de la Deuda del Tesoro y del Estado; séptimo, alza muy acentuada de los fondos públicos y demás efectos valores industriales de renta fija o variable y títulos especulativos, y octavo, estabilización de la moneda»[32].

En cambio, la economía española, según Manuel Azaña, era calamitosa[33]. La República había llegado en el peor momento, y obligaba al Estado a adoptar medidas intervencionistas. En el segundo bienio republicano las leyes se volcaron contra los trabajadores, bajaron los salarios y hubo un permanente acoso a los obreros sindicalizados por parte de los empresarios.

Bien es verdad que algunas leyes trataban de mejorar la hacienda pública como, por ejemplo, la tributación del 30% por los derechos de herencia, pero el problema fundamental radicaba en que las arcas del Estado no se nutrían de los gravámenes a las grandes fortunas como intentó modificar obsesivamente Chapaprieta. Hay que reconocer que tradicionalmente en el estilo tributario mediterráneo ha prevalecido el gravamen sobre el consumo de bienes y servicios frente a los impuestos sobre la renta y los beneficios[34].

En el plano microeconómico, el interés de los préstamos personales nunca superó el 7%. Los salarios durante el primer bienio no bajaron, se incrementaron un 12%, pero este indicador, a juicio de Juan Hernández Andreu[35], no sólo no mejoró la economía del momento sino que contribuyó a aumentar las diferencias entre empleados y parados. Además, conforme los precios bajaban los costes salariales subían, y muchos empresarios se vieron obligados a cerrar sus negocios. Se especula con el elevado número de parados, pero no se indican cifras exactas. Lo que sí es cierto es que las huelgas se incrementaron y los problemas económicos se sumaron a los problemas sociales.

---

30. Instituto Nacional de Estadística, *Anuario Estadístico de España 1858-1997 en* INEbase/ Historia. En línea en http://www.ine.es/prodyser/pubweb/anuarios_mnu.htm [Consulta: 27.09.16].
31. M. Tuñón de Lara, "La Segunda República española", *Cuadernos de Historia 16*, 1, 1995, 6.
32. *Apud* M.C. García Nieto y J.M. Donezar, *La Segunda República. Economía y aparato del Estado 1931-1936*, Madrid; Barcelona, 1974, 214.
33. M. Azaña. *Causas de la guerra de España*, Barcelona, 1986.
34. Fuentes Quintana, *La reforma fiscal… op. cit.*, 314.
35. J. Hernández Andreu, *España y la crisis de 1929, op. cit.*, 15

Pero ¿cómo se encontraba en realidad España? A pesar de los muchos estudios cuantitativos y datos estadísticos aportados recientemente, Jordi Palafox[36] sostiene que aún no contamos con los suficientes para conocer con precisión la evolución de precios, salarios, paro, distribución de la riqueza e incluso renta nacional. En la misma línea se manifiesta el profesor Comín: «hay que comenzar reconociendo que se desconocen tantos y tan relevantes aspectos de la economía española del período de entreguerras que sostener cualquier interpretación contra viento y marea no es lo más sensato»[37].

## 4. El problema de la reforma agraria

Resulta difícil valorar el nivel de bienestar general de entonces sin atender al grueso del potencial de mano de obra de la clase social mayoritaria: el campesinado. La España de principios del siglo XX era uno de los países más atrasados de Europa. Aproximadamente contaba con 23,5 millones de habitantes, de ellos 12 000 familias terratenientes, 40 000 comerciantes, 80 000 grandes empresarios y 4,7 millones de clases medias, de manera que algo más de las tres cuartas partes de la población, aproximadamente, las formaban los asalariados en condiciones miserables. Pero el mayor problema lo constituía la gente del campo que no había buscado mejores oportunidades fuera de su lugar de origen, y que tampoco permitía a la débil burguesía aprovechar este potencial de mano de obra.

Siendo España un país eminentemente agrícola y ganadero, el problema de la propiedad agraria es uno de los asuntos más tratados y debatidos cuando hay que referirse a esta época. La pobreza de jornaleros y pequeños agricultores resultaba vergonzante, sobre todo de Madrid hacia el sur.

Las fuentes de riqueza en gran parte de la geografía nacional dependían, pues, del sector primario. En los años 30 la agricultura estaba poco mecanizada y la producción se repartía por los ámbitos rurales donde, a su vez, se producían grandes desigualdades.

La industria se circunscribía esencialmente a Cataluña y al País Vasco. Madrid era el centro financiero y político del país pero su grado de industrialización era muy limitado.

Los conflictos sociales se produjeron no sólo en las fábricas sino también en el campo, donde la reacción de los terratenientes resultó más virulenta porque atentaban contra el principio de la propiedad privada.

Desde comienzos del siglo XX se venían ocupando ocasionalmente tierras sin labrar sobre todo en el sur del país, generando conflictos entre las masas populares y las fuerzas del orden establecido.

---

36. J. Palafox, *Atraso económico… op. cit.*, 313.
37. En E. Fuentes Quintana (dir.) y F. Comín (coord.), *Economía y economistas… op. cit.*, 1034.

La ocupación por la fuerza de los latifundios improductivos ocasionó cruentas represiones. Uno de los casos más llamativos fue el de Casas Viejas (hoy Benalup de Sidonia) en la provincia de Cádiz.

La reforma republicana del primer bienio intentó remediar estos desajustes pero no se obtuvieron los resultados esperados. Principalmente los motivos hay que buscarlos en el bajo nivel tecnológico de la maquinaria empleada en las labores agrícolas, y en una administración de justicia impotente para resolver de manera rápida los reiterados recursos interpuestos por los terratenientes para blindar sus posesiones.

La reforma agraria del año 1931 inicialmente planteó la expropiación indefinida del usufructo pero no de la propiedad. La *Ley de Reforma Agraria* del 9 de setiembre de 1932 resultó muy difícil de aplicar. Contaba, además, con un límite presupuestario escaso para atender a las indemnizaciones.

Aproximadamente dos millones de jornaleros apenas reunían recursos suficientes para sobrevivir, pero algunos de los decretos para paliar esta lamentable situación se elaboraron de manera precipitada y sin tener a veces en cuenta las consecuencias, como ocurrió con el aparecido el 28 de abril de 1931 por el que se prohibía contratar obreros forasteros. Esto produjo más perjuicios que ventajas puesto que aquellos que acababan las faenas en su pueblo no podían contratarse en otros lugares donde había ofertas de trabajo[38].

La reforma agraria emprendida por la República, como dice Manuel Tuñón de Lara «sólo puso en cuestión las relaciones de trabajo en el campo, nunca las relaciones de producción»[39].

Más adelante, durante la guerra se impondría en algunas regiones el modelo de la colectivización o explotación comunal de los recursos. Este fue un fenómeno que se remonta a los pueblos prerromanos, y en cierta forma ya se había producido durante la alta Edad Media, al menos en Castilla y Aragón, cuajado en las Comunidades de Ciudad o de Villa y Corte[40].

Las colectividades de índole libertaria establecidas en Aragón y Cataluña fueron fundamentales, según Gastón Leval[41], para el avituallamiento de las fuerzas milicianas. Leval considera este sistema más importante que la propia fuerza bruta de las armas para ganar una guerra, pero el éxito económico en este caso no puede ser analizado sino en las circunstancias en las que se originó, es decir, en momentos de guerra y en el entorno de un sistema de producción localista de autoconsumo.

## 5. Actores y actos de la política económica

La idea más extendida acerca de la toma de decisiones de política económica durante los años de paz (1931-1936) es la que la identifica con el continuismo de los patrones del crecimiento hacia adentro y del proteccionismo, surgidos ya desde los tiempos de la formación del Estado español en tiempo de los Reyes Católicos.

---

38. L. Garrido González, "Constitución y reformas socioeconómicas en la España de la II República", *Historia Contemporánea*, 1991, 6, 176.

39. *Apud* L. Garrido González, 1991, 176.

40. A. Carretero y Jiménez, *La personalidad de Castilla en el conjunto de los pueblos hispánicos*, San Sebastián, 1977, 49 ss.

41. G. Leval, *Colectividades libertarias en España*, Madrid, 1977.

Aun cuando nuestro país mantuvo relaciones comerciales con el exterior, en la práctica España no empezó a liberalizar en serio su economía, y comenzó el proceso de apertura al comercio internacional hasta el Plan Nacional de Estabilización Económica de 1959, tomando como base fundamental la paridad de la peseta respecto al dólar.

La tendencia un tanto aislacionista fue apoyada sin duda por las políticas económicas de los sucesivos gobiernos. Este panorama se repitió durante toda la mitad del siglo XX al menos, y con caracteres específicos a partir de 1939. Pero, además, al prevaleciente intervencionismo estatal hay que añadir los escasos conocimientos sobre Economía por parte de la mayoría de los políticos. Una persona, por ejemplo, de alta talla política, como Indalecio Prieto, reconocía su incapacidad para controlar la hacienda pública y se veía obligado a recurrir con frecuencia a Flores de Lemus[42].

En la época republicana una de las pocas excepciones de un ministro de Hacienda con conocimientos de Economía fue Jaume Carner i Romeu. Su labor en el Ministerio fue fundamental. Los únicos dos presupuestos generales que aprobaron las Cortes durante la Segunda República en tiempos de paz fueron los preparados por Carner, el resto fueron prórrogas.

Agustín Viñuales Pardo y Gabriel Franco López fueron catedráticos de Universidad y también desempeñaron la cartera de Hacienda pero su mandato fue muy corto. Ambos duraron tres meses en el cargo, y el primero pasó con más pena que gloria[43].

Los responsables de Hacienda durante la Segunda República en tiempo de paz fueron, por orden cronológico: Indalecio Prieto, Jaume Carner, Agustín Viñuales, Antonio de Lara y Zárate, Manuel Marraco y Ramón, Alfredo Zabala y Zafora, Joaquín Chapaprieta Torregrosa, Manuel Rico Avello, Gabriel Franco y Enrique Ramos[44].

Es pertinente observar que no siempre estos ministros pertenecieron al mismo partido de su presidente de gobierno. Fue el caso de Indalecio Prieto (Partido Socialista Obrero Español) bajo la presidencia de Niceto Alcalá Zamora (Derecha Liberal Republicana), Jaume Carner (Conservador Independiente) bajo Manuel Azaña (Acción Republicana), Joaquín Chapaprieta (Conservador Independiente) bajo Alejandro Lerroux (Partido Republicano Radical), Manuel Rico Avello y García Lañón (independiente) bajo Manuel Portela Valladares (Conservador independiente), Gabriel Franco (Izquierda Republicana) bajo Manuel Azaña (Acción Republicana) y Enrique Ramos (Izquierda Republicana) bajo Santiago Casares Quiroga (Organización Republicana Gallega Autónoma).

---

42. J. M. García Escudero, "La semblanza de los doce ministros de Hacienda de la II República" en R. Calle Saiz, *La hacienda pública en la II República española*, Madrid, 1981, II, 1572. Una cierta idea de nuestra afirmación se puede vislumbrar en las memorias de Prieto (*De mi vida*, México, 1965, v. I) donde apenas se encuentran referencias a la Economía, salvo en las pp. 201-202, en el apartado titulado "El oro español y el cocido madrileño".

43. Calle, *La hacienda pública… op. cit.*, II, 905.

44. Un conciso estudio sobre ellos lo encontramos en García Escudero, "La semblanza de los doce ministros…" *op. cit.*, 1550-1589.

Es más, si echamos una ojeada al cuadro de Comín sobre los ministros de Hacienda entre 1899 y 1936[45], puede apreciarse la prevalencia de una línea más conservadora que liberal entre sus representantes, exceptuando a Antonio de Lara y Zárate, y a Manuel Marraco y Ramón, miembros del Partido Republicano Radical. En el resto de los casos, al tratarse de gobiernos de coalición, los intereses de los diferentes mandatarios ocasionaban a veces ideas contradictorias y actitudes divergentes.

Qué tipo de ósmosis había entre presidentes y ministros es algo que escapa a nuestros propósitos, pero no siempre se producían las simpatías necesarias para una labor de equipo. En sus *Memorias políticas y de guerra*[46], Manuel Azaña, por ejemplo, llega a burlarse de ciertas actitudes de Niceto Alcalá Zamora. Asimismo Joaquín Chapaprieta desmiente en sus *Memorias*[47] su supuesta amistad con Alcalá Zamora así como las presiones a las que con frecuencia se veía sometido para formar un equipo de trabajo.

La influencia de los economistas dedicados a la docencia es algo que también habría que considerar. A través de sus publicaciones (como hicieran los arbitristas de siglos anteriores), los políticos podían obtener informaciones técnicas para la toma de decisiones.

Uno de los economistas más lúcidos del momento fue el profesor Germán Bernácer, seguidor del liberalismo clásico, y partidario de la estabilización de la moneda, pero no de su integración en el patrón oro, en contra de las ideas de Flores de Lemus. Según el profesor Hernández Andreu[48], hay razones suficientes (idea que nos parece discutible) para sospechar que las ideas de Bernácer calaron en la política económica de la República.

\* \* \*

Parece haber un consenso general acerca de la apropiada labor de los ministros de Hacienda: Indalecio Prieto, Jaume Carner y Joaquín Chapaprieta. Sus decisiones fueron fundamentales para atajar la constante tendencia inflacionista heredada de la Dictadura y conseguir la estabilización de la moneda. De hecho, durante los años de la República, excepto en 1934, la Hacienda Pública no fue deficitaria[49].

---

45. F. Comín, P. Martín Aceña y M. Martorell Linares (eds.), *La Hacienda desde sus ministros. Del 98 a la guerra civil*, Zaragoza, 2000, 15.

46. M. Azaña, *Memorias políticas y de guerra*, Barcelona, 1981.

47. J. Chapaprieta Torregrosa, *La paz fue posible. Memorias de un político*, Barcelona, 1971.

48. J. Hernández Andreu, *Pensamiento y economía monetaria en España durante la Segunda República*. Documento de trabajo 9126, Biblioteca de la Facultad de CC. Económicas y Empresariales de la Universidad Complutense de Madrid, s.a.

49. J. Hernández Andreu, "Análisis de gastos públicos discrecionales: La política presupuestaria de la Segunda República española", *Revista de Historia Económica*, 1993, XI, 1, 93.

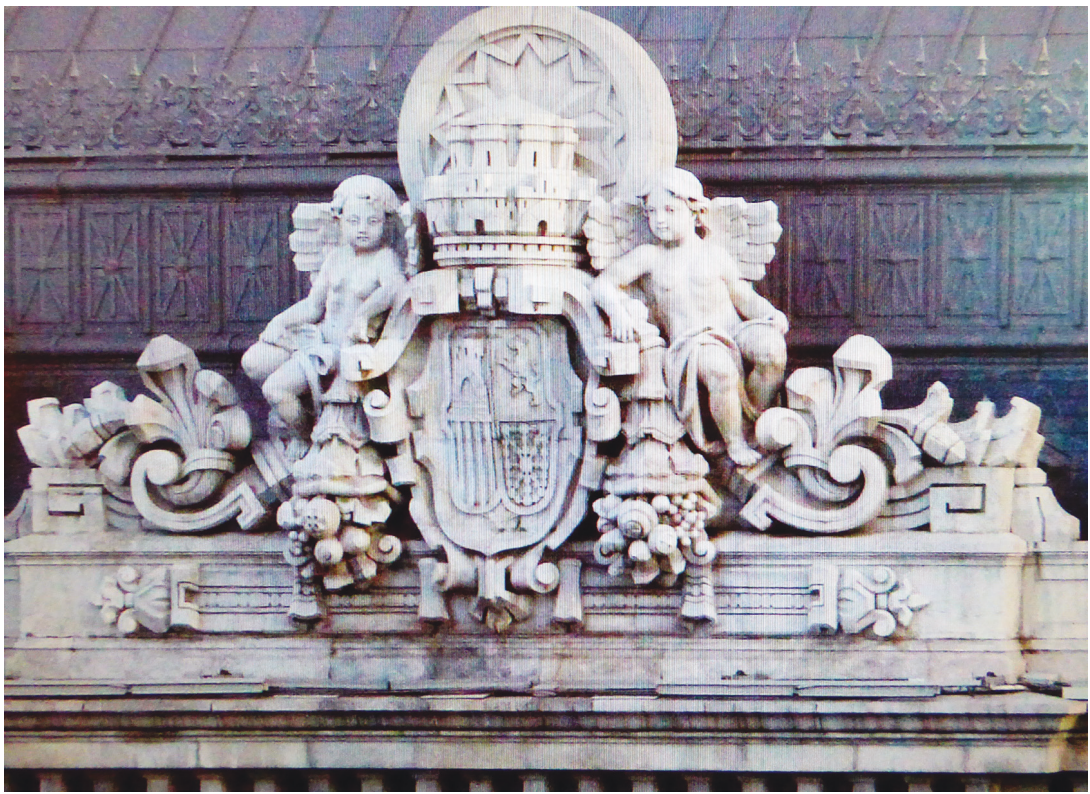

**Escudo de la Segunda República en el edificio del Banco de España.
(Fotografía de los autores)**

Prieto vio imprescindible la intervención del gobierno en el Banco de España[50] para determinar sobre todo las variaciones en los tipos de descuento e interés. Carner y Chapaprieta fueron los impulsores de la necesaria reforma tributaria y de medidas restrictivas como la reducción de las importaciones, el equilibrio del presupuesto y la remodelación de la burocracia. Esta política deflacionista no provocó la contracción de la demanda interna como en alguna ocasión se ha dicho[51] y pretendía optimizar la distribución de la renta.

No obstante, a pesar de estas decisiones pretendidamente correctoras, «El poder económico siguió en manos de la oligarquía financiera y terrateniente: el bloque dominante había perdido la dirección del Estado pero conservaba intactas sus poderes económicos y sociales»[52].

---

50. Su nacionalización no se produciría hasta 1962.

51. N. García Santos y P. Martín-Aceña, "El comportamiento del gasto público en España durante la Segunda República, 1931-1935", *Revista de Historia Económica*, 1990, 2, 398.

52. García Nieto y Donézar, *La Segunda República. Economía… op. cit.*, 28-29.

Con el fin de mejorar los servicios públicos, los gobiernos de la Segunda República subieron los impuestos como única vía para equilibrar el presupuesto, habida cuenta de las trabas de las clases privilegiadas para una reforma profunda del sistema tributario. El aumento de la presión fiscal respecto de la Dictadura fue mayor, como afirma Ramón Tamames[53], en cambio la recaudación decreció; en consecuencia se puede percibir que el fenómeno de la defraudación solía ser moneda corriente.

En un ambiente como este, de constantes tiranteces e inseguridad, la producción decayó. No obstante, hay que dejar bien claro que las dificultades económicas no se debieron a la llegada de la Segunda República porque ya se producían en la dictadura de Primo de Rivera.

Entre 1928 y 1932 la peseta se vio sometida a sucesivas devaluaciones respecto al patrón oro. Pero a partir de 1933 se hizo necesaria una revaluación de la moneda porque, como señala Juan Hernández Andreu, la tendencia anterior: «1) Encarecía las importaciones; 2) no beneficiaba a las exportaciones, ya que los precios mundiales caen desde 1929; y 3) había que evitar efectos inflacionistas en el interior»[54].

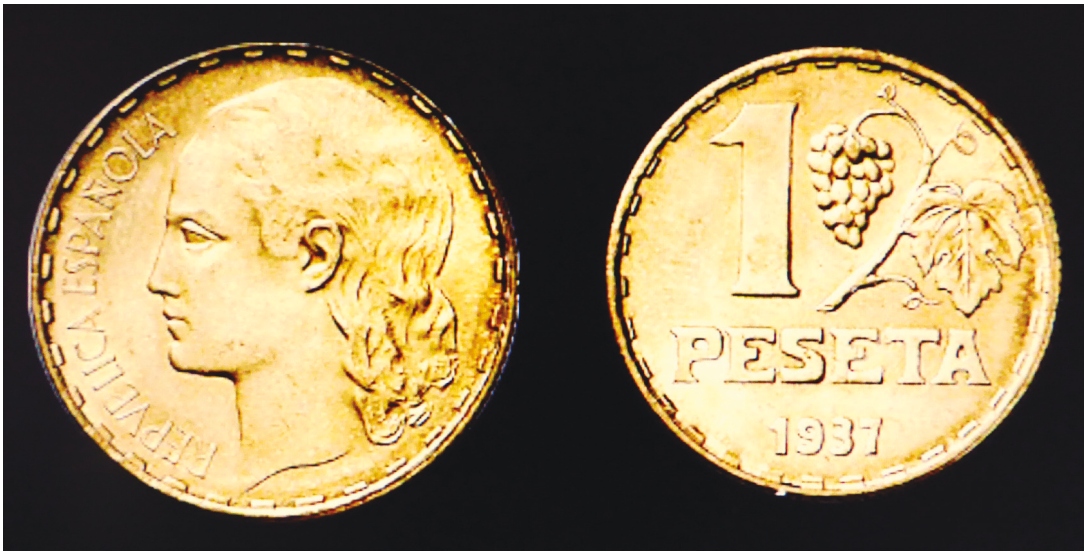

**2. Moneda de latón, conocida popularmente como la rubia por el tono de su colorido. En el anverso figura un rostro de mujer con clara alusión a la República. (Fotografía de los autores)**

Desde el punto de vista de la hacienda pública, las decisiones gubernamentales en general fueron acertadas. Pero el problema fundamental no era ese, sino el de la distribución de la riqueza y opacidad en la justificación del gasto. En un apartado tan simbólico como el de las ayudas eclesiásticas, por ejemplo, podemos hacernos una ligera idea acerca de las ventajas de la iglesia católica, para sus 35 000 miembros del clero secular, a partir de la tesis doctoral de Manuel Pascual Rodríguez *Dotación de la Iglesia española en la Segunda República*[55].

---

53. R. Tamames, *La República – La era de Franco*, Madrid, 1975, *passim* 125-131.
54. J. Hernández Andreu, *doc. trab. 9126… op. cit.*, 2.

Parece apropiado señalar que la Iglesia nunca dejó de percibir ayuda económica durante la Segunda República porque a pesar de haberse tomado la decisión de que un país laico no debía apoyar económicamente a las instituciones religiosas, la idea de los gobiernos del primer bienio fue la de ir reduciendo este gasto paulatinamente hasta su desaparición durante los cuatro años venideros a razón de una disminución del 25% anual; decisión que no se llegó a consumar porque fue derogada por los gobiernos conservadores.

\* \* \*

¿Cómo se deben interpretar los datos apuntados anteriormente? Si es cierto que el éxito o fracaso de un gobierno depende fundamentalmente de la política económica, en el caso de la Segunda República española se puede afirmar que las medidas económicas estuvieron a merced de los ideales políticos o, dicho de otro modo, se prestó más atención a eliminar las medidas opresoras del pasado que a sentar un sistema económico sólido.

A pesar del retraso económico que sobrellevaba España en relación a los países centroeuropeos, los problemas económicos no fueron los temas prioritarios, pero aludir a este hecho como factor desencadenante de disturbios sociales, y finalmente de la guerra civil, es una interpretación sesgada, propia de historiadores conservadores[56].

Tanto los mandatarios durante el bienio azañista, con gobiernos de centro-izquierda, como los gobiernos conservadores posteriores, y como los líderes del Frente Popular pusieron mucho empeño en solucionar los problemas sociales más apremiantes, aunque fuera desde perspectivas diferentes.

Su objetivo fundamental fue el control de una sociedad resquebrajada e injusta. A partir de 1934, con el triunfo electoral de las fuerzas conservadoras en el Parlamento, las diferencias entre pobres y ricos se agudizaron y la violencia prendió en las calles, propiciada por grupos revolucionarios de izquierdas y de derechas. La revolución de octubre fue el hecho más trágico. Ocupados en los desórdenes sociales, la economía ocupó un segundo plano entre las preocupaciones de los políticos.

La inseguridad social hizo que los capitales huyeran. Quienes sacaron el dinero a Estados Unidos y Reino Unido se llevaron enormes sorpresas con la gran devaluación de sus monedas, y no fueron menores que los contratiempos experimentados en Francia, Holanda o Suiza. Ante estos imprevistos descalabros, muchos capitales retornaron. La sol-vencia del Tesoro parecía evidente, habida cuenta del éxito que supuso la deuda pública o los empréstitos de 1932 y de 1936, cuya demanda superó a la oferta.

---

55. M. Pascual Rodríguez, *Dotación de la Iglesia española en la Segunda República*, tesis doctoral. Madrid. Universidad Complutense. Facultad de Derecho, 1993, texto completo en http://eprints.ucm.es/tesis/19911996/S/0/S0014401.pdf [Consulta 06.01.15].

56. Cfr. S. V. Florensa Palau, "Economía y política económica de la II República. Una nota de síntesis", *Arbor Ciencia Pensamiento y Cultura*, 1981, CIX, 426, 257.

## 6. Apéndice. El futuro inmediato

La creencia de que la Segunda República española fue un sistema político inadecuado para una población no preparada para asumirlo nos parece una información divulgativa tendenciosa. El problema no era el sistema republicano sino su sociedad poco socializada y con graves problemas de convivencia debidos principalmente a las diferencias de clase.

Un enfrentamiento social que termine en una guerra civil es el resultado de muchas variables. Aquí hemos elegido el aspecto económico pero indudablemente hay muchos otros motivos que la ocasionan. Por lo que a la economía se refiere, podemos asegurar que se emplearon medidas similares a las de otros países europeos para sacar al país del atraso en el que se encontraba.

La política económica de la Segunda República, a pesar de algunas veleidades, impuestas por los gobernantes de turno, supo paliar los efectos de la crisis internacional de manera airosa. La situación cambiaría con los trágicos acontecimientos que se avecinaban y aquí sí que los móviles económicos resultaron determinantes para la victoria o para la derrota.

Las medidas económicas durante la contienda se ajustarían a lo que es una economía de guerra; es decir, el intervencionismo gubernamental que por ambos lados sería intenso, con la salvedad apuntada más arriba sobre las colectivizaciones.

Algunas de las preguntas más relevantes acerca de la actividad económica se refieren a la gestión y distribución de los recursos materiales por parte de los contendientes, y a si la economía fue determinante para ganar o perder la guerra.

Por lo que concierne a la primera cuestión, contamos, entre otros, con dos clarificadores estudios, el del profesor Viñas[57], quien descubrió un documento en la Fundación Universitaria Española en el que se detalla la compra a Italia de material para preparar una guerra (no un golpe de Estado), y la obra de conjunto editada por Pablo Martín Aceña y Elena Martínez Ruiz[58], según los cuales el bando rebelde contó con mejor gestión de recursos y mayor ayuda exterior que el bando oficialista.

En el segundo caso creemos que la República tenía perdida la guerra desde el principio, coincidiendo en parte con el vaticinio de Indalecio Prieto[59] de que el sistema republicano estaba herido de muerte nada más comenzar la rebelión castrense. La falta de apoyo de las democracias exteriores temerosas del triunfo de una revolución social española de izquierdas que contribuyera a desestabilizarlas, la ayuda rusa que resultó exigua a partir de 1937 y el desplazamiento del Gobierno de Madrid a Valencia a los tres meses de haber comenzado el conflicto eran indicios de la imposibilidad de vencer a los sublevados.

---

57. Á. Viñas, *La República en guerra. Contra Franco, Hitler, Mussolini y la hostilidad británica*, Barcelona, 2012.
58. P. Martín Aceña y E. Martínez Ruiz (eds.), *La economía de la guerra civil*, Madrid, 2006.
59. I. Prieto, *Convulsiones de España*, México D.F, 1967-1969, 3 vols.

## 7. Conclusiones

La Segunda República española, metafóricamente *La Niña Bonita*, fue ante todo un proyecto de futuro y no el fracaso de un sistema político y económico, puesto que en tan corto espacio temporal no tuvo tiempo de desarrollarse.

En el fondo, la economía jugó un papel trascendental al que no se le ha prestado toda la atención debida. La economía fue determinante en los avatares políticos que acabarían en los luctuosos acontecimientos del 36, en el sentido de que las fuerzas reaccionarias tenían un proyecto de futuro consistente en la recuperación de sus privilegios. Así, la guerra civil, lejos de parecer el golpe militar de unos descontentos con los desórdenes sociales, fue más bien un programa de amplio espectro, cuidadosamente preparado con grandes inversiones en material bélico pesado, como recientemente ha demostrado el profesor Viñas[60].

Más que la defensa de los valores tradicionales, predominó el espíritu de depredación característico de las clases altas y de algunas bajas, como el caso del general Franco, quien recién nombrado jefe de los ejércitos, ejerció un ambicioso plan de enriquecimiento personal[61]

La bibliografía sobre el período de la Segunda República española resulta tan controvertida como la propia historia, pero al menos podemos convenir que la política económica española durante la Segunda República tuvo un acentuado carácter intervencionista y no avanzó en la liberalización exterior a pesar de algunos intentos aislados. Algo, por cierto, difícil en un período en que los países se protegían de la propagación de la crisis económica internacional a través de controles comerciales y devaluaciones competitivas.

La necesidad de cambios estructurales en la sociedad, destinados a reparar injusticias sociales se manifiesta primordialmente en la *Constitución* y otras disposiciones legislativas. Se aprobaron leyes, en algunos casos, avanzadas para su tiempo, pero en particular, las referentes a la tenencia de la tierra provocaron el descontento de la oligarquía y de las clases conservadores que se sintieron amenazadas y promovieron un movimiento reaccionario que fomentó las desigualdades entre clases sociales y puso trabas al desarrollo económico.

Desde la perspectiva macroeconómica, los intentos legislativos durante el primer bienio por mejorar la vida de los más desfavorecidos tuvieron dificultades para ponerlos en práctica y sirvieron poco para reparar las abrumadoras miserias de la mayor parte de la población en medio de un sistema democrático todavía poco asentado. Durante el segundo bienio, los gobiernos conservadores iniciaron una contrarreforma que se tradujo principalmente en la recuperación de privilegios, recorte de libertades y aumento de las desigualdades entre ricos y pobres.

Los gobernantes, en su mayoría, prestaron más atención a los problemas sociales que a los económicos porque estos resultaban menos urgentes para la estabilización del país. La República tuvo que hacer frente principalmente al anarquismo, al conservadurismo monárquico, a la iglesia católica y gran parte del ejército, y a ciertos grupos de presión (ra-mificaciones de partidos, sindicatos,...). Pero el atraso económico no condujo ni determinó el estallido de la guerra civil, puesto que el momento más crítico para la economía española se produjo en 1930 (en tiempos aún de la dictadura previa), y en los años siguientes hubo una relativa recuperación.

---

60. M. Amorós, *75 años después: las claves de la Guerra Civil Española: conversación con Ángel Viñas*, Barcelona, 2014.

61. Cfr. Á. Viñas. *La otra cara del caudillo*. Barcelona, 2015.

En el ámbito económico, es difícil establecer un sistema ideal en un lugar cerrado. Las islas financieras no existen, con excepción de los paraísos fiscales. En este sentido, la política económica de los gobiernos de la Segunda República española se ajustó a las costumbres del comercio internacional dentro de la escasa proyección de la economía española hacia el exterior, y en tales decisiones, prevalecieron los criterios técnicos por encima de los sentimientos populares.

Las desproporcionadas diferencias en la distribución de la renta durante la Segunda República española, sin ser el motivo principal que desatara la guerra civil, constituyen, sin embargo, una preocupación que invita a realizar estudios comparativos con la realidad actual.

La ventajosa neutralidad española en las dos guerras mundiales se vio desbaratada por el hecho de la guerra civil que impidió una vez más la oportunidad de que España figurara como una de las principales potencias económicas del planeta.